\documentclass[journal]{IEEEtran}

\usepackage{CJKutf8}

\usepackage{graphicx}
\usepackage{amsmath}
\usepackage{cite}

\begin{document}

\title{Adjustable AprilTags For Identity Secured Tasks}

\author{Hao Li  
\thanks{Hao Li, Dept. Automation, SJTU, Shanghai, 200240, China. CTO of Qingfei.AI. e-mail: haoli@sjtu.edu.cn}
}

\maketitle

\begin{abstract}
Special tags such as AprilTags that facilitate image processing and pattern recognition are useful in practical applications. In close and private environments, identity security is unlikely to be an issue because all involved AprilTags can be completely regulated. However, in open and public environments, identity security is no longer an issue that can be neglected. To handle potential harm caused by adversarial attacks, this note advocates utilization of adjustable AprilTags instead of fixed ones.
\end{abstract}

\IEEEpeerreviewmaketitle

\section{Introduction}

Special tags that facilitate image processing and pattern recognition are useful in practical applications. For example, the AprilTags \cite{Olson2012} \cite{WangJ2016} proposed by the April Robotics Laboratory at the University of Michigan in recent years are a representative kind of special tags that can benefit vision based intelligent systems.

In close and private environments such as in the context of AprilTag map based warehouse robot visual navigation \cite{Li2023AprilTagNavigation}, identity security is unlikely to be an issue because all involved AprilTags for robots can be completely regulated in any desirable way.

However, in open and public environments such as in the context of autonomous public bus oriented vehicle-road cooperative localization \cite{Li2025VTT}, identity security is no longer an issue that can be neglected. What if adversarial attackers fake AprilTags that possess special meanings in the applications and maliciously cause troubles. To handle potential harm caused by adversarial attacks, this note advocates utilization of adjustable AprilTags instead of fixed ones.

\section{AprilTag basics}

An AprilTag is a square-shape tag. It consists of a white outer boundary, a black inner boundary, and an internal matrix of black-white cells. For the AprilTag, its internal matrix of black-white cells can encode its identity. AprilTags just of a moderate internal matrix size are already enough to encode a large amount of distinct identities. For example, consider the kind of AprilTags of a $6$-by-$6$ internal matrix. Theoretically speaking, all such AprilTags allow encoding of $2^{6 \times 6} = 68719476736$ distinct identities. In practice, it is unnecessary to use all these AprilTags but only a small portion of them. Using only a small portion of all possible AprilTags has another merit, namely that the AprilTags actually adopted can be designed to have large mutual Hamming distances  \cite{Li2025GFCV_en} among them. The larger the mutual Hamming distances among AprilTags are, the less likely they will be confused and hence the more robust the AprilTag recognition will be.

\begin{figure}[h!]
\begin{center}
\includegraphics[width=0.99\columnwidth]{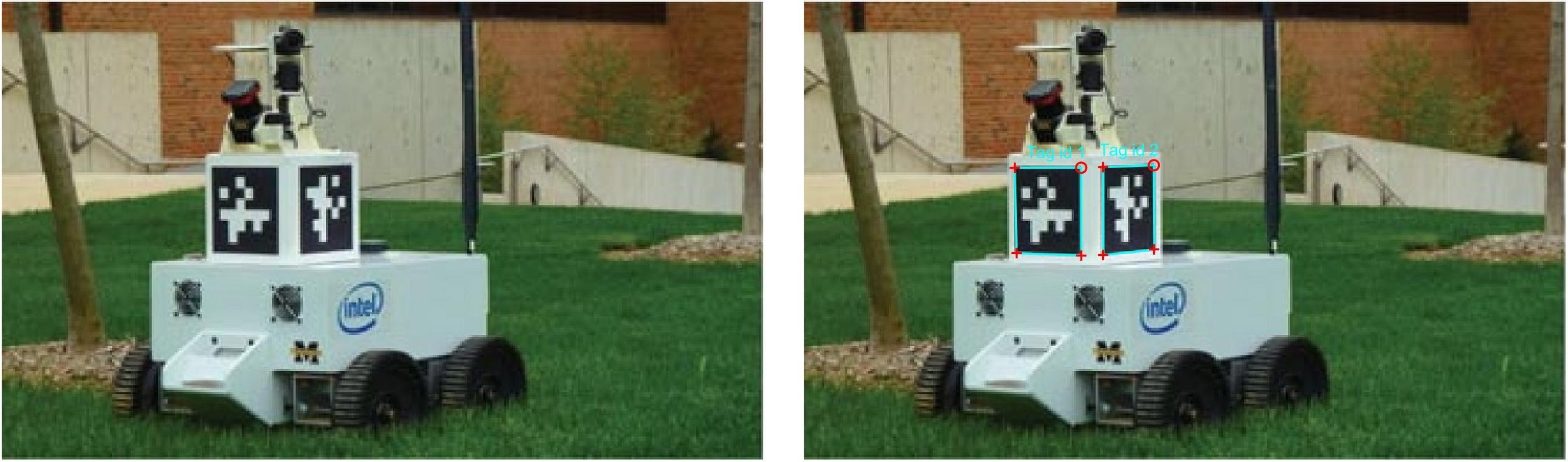}
\end{center}
\caption{One robot carrying AprilTags: (top) original image (figure from \cite{Olson2012}); (bottom) AprilTag detection}
\label{fig:apriltag_one_org+detection}
\end{figure}

AprilTags can be attached to mobile robots to distinguish their identities, which can largely facilitate data association in practical applications. For example, a mobile robot that carries distinct AprilTags on its sides is illustrated in Figure \ref{fig:apriltag_one_org+detection}.

\begin{figure}[h!]
\begin{center}
\includegraphics[width=0.99\columnwidth]{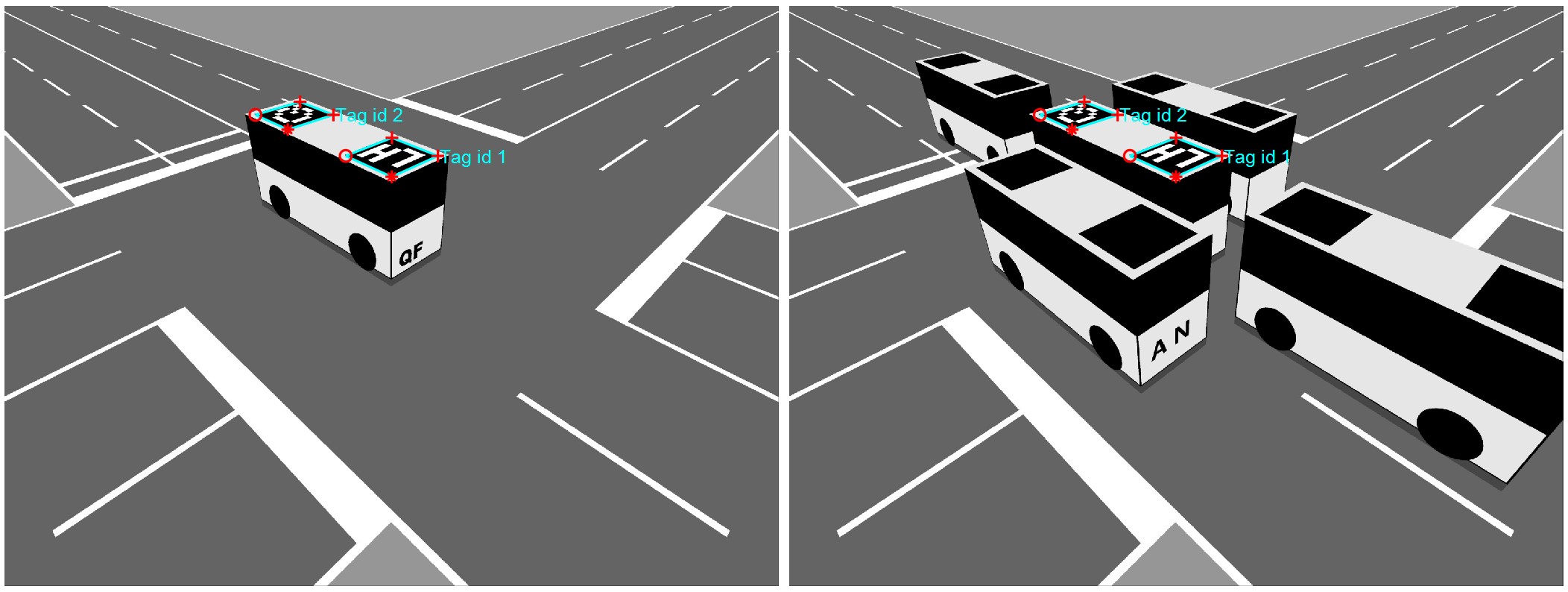}
\end{center}
\caption{AprilTag detection (including tag control point extraction) in the context of autonomous public bus applications}
\label{fig:apriltag_detection}
\end{figure}

For another example, an autonomous public bus simulated with physical configurations rooted in practical applications, which installs AprilTags on its top, is illustrated in Figure \ref{fig:apriltag_detection}. Thanks to the vehicle top tag, the autonomous public bus can be detected and localized reliably and accurately by perceptive roadside units (RSUs) \cite{Li2025VTT} --- It is worth noting that the application scenario demonstrated by the right sub-figure of Figure \ref{fig:apriltag_detection} is especially difficult to general vehicle detection (even with state-of-the-art methods) \cite{WangZ2023} \cite{LiangL2024}, as the target autonomous public bus is surrounded by big vehicles and hence suffers from severe view occlusion.

On one hand, special tags such as AprilTags are usually so distinct and unique in the environment perceived by involved vision systems that normally neither false negative (missed detection) nor false positive (wrong detection) has any chance to exist. Besides, the special tags are highly structured and geometrically regular, so tag detection and localization are completely based on rigorous logic rules and geometry principles. In other words, they are completely explainable, which further accounts for reliability of tag detection and localization.

On the other hand, when adversarial attackers intentionally fake AprilTags intended for certain applications, the special tags are then no longer unique in the environment and hence confusion and even more severe harm can be caused.

\section{Adjustable AprilTags for identity secured applications: Example of Identity secured vehicle-road cooperation}

To handle potential harm caused by adversarial attackers that fake target AprilTags for malicious purpose, this note advocates utilization of adjustable AprilTags that can be realized by simple electronic screen or lighting devices.

To clarify the idea of adjustable AprilTags, take the application of \textit{vehicle top tag assisted vehicle-road cooperative localization} \cite{Li2025VTT} or for short \textit{vehicle-road cooperative localization} as example. The practice of vehicle-road cooperative localization stems from the need of vehicle localization in complicated application scenarios like intersections, which can be especially risky because they can be crowded by dynamic objects coming from and going in various directions. Besides, there is not any lane mark for guidance at intersections.

Instead of single vehicle operation that has inherent limitations, vehicle-road cooperative localization takes advantage of the V2X (vehicle-to-everything) strategy, or more specifically, of cooperation between the vehicle and other intelligent entities in traffic environments. Compared with multi-vehicle cooperative localization \cite{Li2013ITSMag, Li2014TITS, Li2024TITS} which has difficulty rather at policy and administration levels, vehicle-road cooperative localization would be more sound, as intersections are already equipped with perceptive roadside units (RSUs) thanks to rapidly-developing ``vehicle-road-cloud'' projects.

\subsection{Basic methodology of vehicle top tag assisted vehicle-road cooperative localization} 

For vehicle top tag assisted localization, when the autonomous public bus approaches an intersection, it communicates with visual RSUs at the intersection and inform them to initiate the process of vehicle top tag detection and localization. 

Upon initiating the process, visual RSUs try to detect vehicle top tags first, then perform vehicle top tag localization (i.e. vehicle top tag pose estimation) according to detected results, and next share pose estimates with the autonomous public bus. So the autonomous public bus can finally localize itself thanks to pose estimates shared by visual RSUs. 

\begin{figure}[h!]
\begin{center}
\includegraphics[width=0.99\columnwidth]{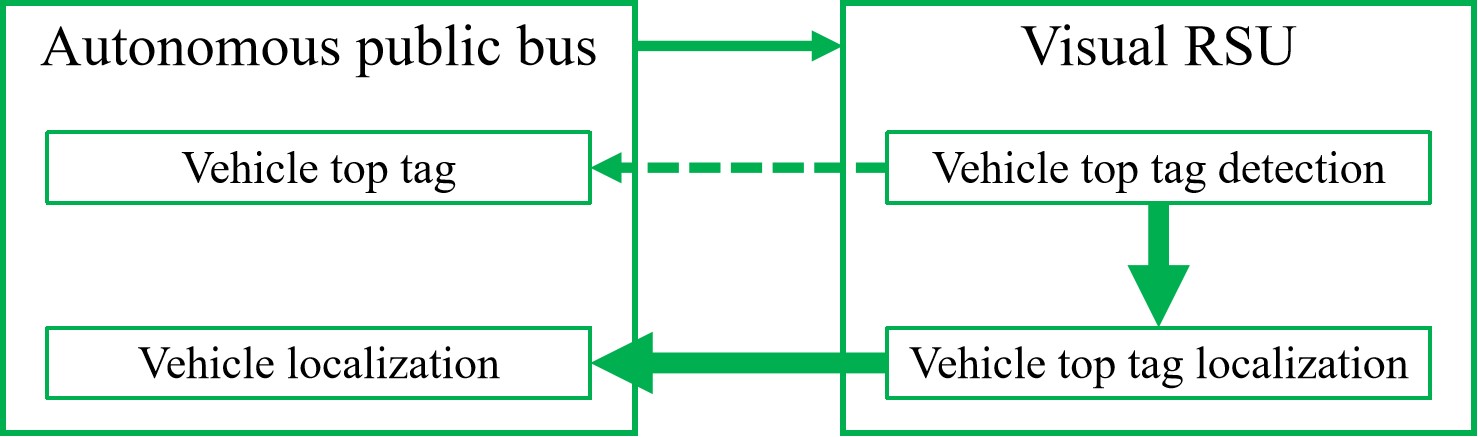}
\end{center}
\caption{Vehicle top tag assisted vehicle-road cooperative localization \cite{Li2025VTT}}
\label{fig:vehicle_road_cooperative_localization}
\end{figure}

When the autonomous public bus leaves the intersection, it informs visual RSUs there to close the process of vehicle top tag detection and localization (otherwise, visual RSUs will perform the process in vain and waste computational consumption). The working mechanism of the proposed methodology of vehicle top tag assisted localization is demonstrated in Figure \ref{fig:vehicle_road_cooperative_localization}.

\subsection{Identity secured version}

When there is no abnormality during cooperation between the autonomous public bus and visual RSUs, vehicle-road cooperative localization is realized following above basic methodology demonstrated in Figure \ref{fig:vehicle_road_cooperative_localization}.
 
On the other hand, when there are adversarial attackers that fake AprilTags installed on the top of the autonomous public bus, visual RSUs can communicate the status of AprilTag confusion with the autonomous public bus which then will be aware of existence of adversarial attackers.

Upon being informed of potential adversarial attackers, the autonomous public bus adjusts the pattern displayed on the electronic AprilTag device and communicates the identity code of the new AprilTag pattern with visual RSUs which then will know which traffic entity carrying AprilTags is the true ``client'' namely the autonomous public bus whereas which others also carrying AprilTags are fake ones namely adversarial attackers.

Cunning adversarial attackers may pay close attention to the autonomous public bus and follow it to adjust their fake AprilTags accordingly. For sake of handling such case, the autonomous public bus and visual RSUs can synchronize with each other and appoint the moment of their simultaneous ``actions'' namely AprilTag adjustment and image capturing. 

\section{Conclusion}  

Following the basic methodology of vehicle top tag assisted vehicle-road cooperative localization, utilization of adjustable AprilTags instead of fixed ones is advocated to handle potential adversarial attackers that fake target AprilTags for malicious purpose.

\bibliographystyle{IEEEtran}
\bibliography{LI_Ref}

\begin{thebibliography}{10}
\providecommand{\url}[1]{#1}
\csname url@samestyle\endcsname
\providecommand{\newblock}{\relax}
\providecommand{\bibinfo}[2]{#2}
\providecommand{\BIBentrySTDinterwordspacing}{\spaceskip=0pt\relax}
\providecommand{\BIBentryALTinterwordstretchfactor}{4}
\providecommand{\BIBentryALTinterwordspacing}{\spaceskip=\fontdimen2\font plus
\BIBentryALTinterwordstretchfactor\fontdimen3\font minus
  \fontdimen4\font\relax}
\providecommand{\BIBforeignlanguage}[2]{{%
\expandafter\ifx\csname l@#1\endcsname\relax
\typeout{** WARNING: IEEEtran.bst: No hyphenation pattern has been}%
\typeout{** loaded for the language `#1'. Using the pattern for}%
\typeout{** the default language instead.}%
\else
\language=\csname l@#1\endcsname
\fi
#2}}
\providecommand{\BIBdecl}{\relax}
\BIBdecl

\bibitem{Olson2012}
E.~Olson, J.~Strom, R.~Morton, A.~Richardson, P.~Ranganathan, R.~Goeddel,
  M.~Bulic, J.~Crossman, and B.~Marinier, ``Progress toward multi-robot
  reconnaissance and the magic 2010 competition,'' \emph{Journal of Field
  Robotics}, vol.~29, no.~5, p. 2012, 762-792.

\bibitem{WangJ2016}
J.~Wang and E.~Olson, ``Apriltag 2: efficient and robust fiducial detection,''
  in \emph{IEEE/RSJ International Conference on Intelligent Robots and
  Systems}, 2016, pp. 4193--4198.

\bibitem{Li2023AprilTagNavigation}
S.~Fang, Y.~Li, and H.~Li, ``Split covariance intersection filter based visual
  localization with accurate apriltag map for warehouse robot navigation,''
  \emph{arXiv}, 2023.

\bibitem{Li2025VTT}
H.~Li, B.~Liu, and L.~Wang, ``Vehicle top tag assisted vehicle-road cooperative
  localization for autonomous public buses,'' \emph{arXiv}, 2025.

\bibitem{Li2025GFCV_en}
\begin{CJK}{UTF8}{gbsn}李颢\end{CJK},
  \emph{\begin{CJK}{UTF8}{gbsn}计算机视觉几何基础（英文版）\end{CJK}}.\hskip
  1em plus 0.5em minus 0.4em\relax
  \begin{CJK}{UTF8}{gbsn}上海交通大学出版社\end{CJK}, 2025.

\bibitem{WangZ2023}
Z.~Wang, J.~Zhan, C.~Duan, X.~Guan, P.~Lu, and K.~Yang, ``A review of vehicle
  detection techniques for intelligent vehicles,'' \emph{IEEE Transactions on
  Neural Networks and Learning Systems}, vol.~34, no.~8, pp. 3811--3831, 2023.

\bibitem{LiangL2024}
L.~Liang, H.~Ma, L.~Zhao, X.~Xie, C.~Hua, M.~Zhang, and Y.~Zhang, ``Vehicle
  detection algorithms for autonomous driving: a review,'' \emph{Sensors},
  vol.~24, no.~10, pp. 3088(1--38), 2024.

\bibitem{Li2013ITSMag}
H.~Li and F.~Nashashibi, ``Cooperative multi-vehicle localization using split
  covariance intersection filter,'' \emph{IEEE Intelligent Transportation
  Systems Magazine}, vol.~5, no.~2, pp. 33--44, 2013.

\bibitem{Li2014TITS}
H.~Li, M.~Tsukada, F.~Nashashibi, and M.~Parent, ``Multivehicle cooperative
  local mapping: a methodology based on occupancy grid map merging,''
  \emph{IEEE Transactions on Intelligent Transportation Systems}, vol.~15,
  no.~5, pp. 2089--2100, 2014.

\bibitem{Li2024TITS}
S.~Fang and H.~Li, ``Multi-vehicle cooperative simultaneous {L}i{DAR} {SLAM}
  and object tracking in dynamic environments,'' \emph{IEEE Transactions on
  Intelligent Transportation Systems}, vol.~25, no.~9, pp. 11\,411--11\,421,
  2024.

\end{thebibliography}

\end{document}